\documentclass[prd,aps,superscriptaddress,tightenlines,nofootinbib,floatfix,preprintnumbers,11pt,longbibliography]{revtex4-1}
\usepackage{epsfig}
\usepackage{amsfonts}
\usepackage{amsmath}
\usepackage{amssymb}
\usepackage{dsfont}
\usepackage{hyperref}
\hypersetup{
    colorlinks=true,       % false: boxed links; true: colored links
    linkcolor=blue,          % color of internal links
    citecolor=blue,        % color of links to bibliography
    filecolor=blue,      % color of file links
    urlcolor=blue           % color of external links
}
\usepackage{xcolor}
\usepackage{color}

\renewcommand{\vec}[1]{{\mathbf #1}}

\newcommand{\g}{\gamma}

\newcommand{\slashpi}{\slash\hspace{-0.5em}\pi}

\newcommand{\mpi}{m_{\pi}}

\newcommand{\vL}{\ensuremath{\mathcal{L}}}

\newcommand{\sq}{^{2}}

\newcommand{\dslash}[1]{#1 \llap{/\kern-0.5pt}}
\newcommand{\Dslash}[1]{#1 \llap{/\kern+1.5pt}}
\newcommand{\DDslash}[1]{#1 \llap{/\kern+2.3pt}}
\newcommand{\dslashh}[1]{#1 \llap{/\kern+1pt}}

\newcommand{\boldtau}{\mbox{\boldmath $\tau$}}
\newcommand{\boldpi}{\mbox{\boldmath $\pi$}}

\newcommand{\boldsigma}{\mbox{\boldmath $\sigma$}}

\newcommand{\bea}{\begin{eqnarray}}
\newcommand{\eea}{\end{eqnarray}}
\newcommand{\be}{\begin{equation}}
\newcommand{\ee}{\end{equation}}
\newcommand{\bma}{\begin{pmatrix}}
\newcommand{\ema}{\end{pmatrix}}
\newcommand{\nn}{\nonumber}

\newcommand{\NLDBD}{$0 \nu \beta \beta$}
\newcommand{\qhat}{\hat{q}}

\begin{document}
\preprint{LA-UR-17-28401}

\title{
\vspace*{0.5cm}
\Large
Neutrinoless double beta decay 
in  effective field theory: \\
the light Majorana neutrino exchange mechanism
\vspace*{.5cm}
}

\author{Vincenzo Cirigliano}
\affiliation{Theoretical Division, Los Alamos National Laboratory, Los Alamos, NM 87545, USA}

\author{Wouter Dekens}
\affiliation{Theoretical Division, Los Alamos National Laboratory, Los Alamos, NM 87545, USA}
\affiliation{New Mexico Consortium, Los Alamos Research Park, Los Alamos, NM 87544, USA}

\author{Emanuele Mereghetti}
\affiliation{Theoretical Division, Los Alamos National Laboratory, Los Alamos, NM 87545, USA}

\author{Andr\'{e} Walker-Loud}
\affiliation{Nuclear Science Division, Lawrence Berkeley National Laboratory, Berkeley, CA 94720, USA}

\begin{abstract}

\vspace*{.75cm}

We present the first  chiral effective theory  derivation of
the neutrinoless double beta-decay $nn\rightarrow pp$ potential induced by light Majorana neutrino exchange.
The effective-field-theory framework has allowed us to identify and parameterize 
short- and long-range  contributions previously missed in the literature.
These contributions can not be absorbed into parameterizations of the single nucleon form factors.
Starting from the quark and gluon level, we perform the matching onto chiral effective field theory and subsequently onto the nuclear potential. 
To derive the nuclear potential mediating neutrinoless double beta-decay, the hard, soft and potential neutrino modes must be  
integrated out.  
This is performed through next-to-next-to-leading order in the chiral power counting, 
in both the Weinberg and pionless schemes.
At next-to-next-to-leading order, the amplitude  receives  additional contributions from the exchange of ultrasoft neutrinos, 
which   can be expressed in terms of nuclear matrix elements 
of the weak current and excitation energies of the intermediate nucleus. These quantities also control the 
two-neutrino double beta-decay amplitude. 
Finally, we outline  strategies to determine the low-energy constants that appear in the  potentials, by relating them to electromagnetic couplings and/or by matching to lattice QCD calculations.

\end{abstract}
\maketitle

\section{Introduction} 
The observation of  neutrinoless double beta decay  (\NLDBD) would be direct evidence of 
lepton number violation (LNV)  beyond the Standard Model (SM),
demonstrating that neutrinos are Majorana fermions~\cite{Schechter:1981bd},  shedding light on 
the mechanism of neutrino mass generation, and probing  a key ingredient (LNV)  for  generating the matter-antimatter asymmetry in the universe via 
``leptogenesis''~\cite{Davidson:2008bu}. 
The current experimental  limits on the half-lives are  quite  
impressive~\cite{KamLAND-Zen:2016pfg, Alfonso:2015wka,Albert:2014awa,Agostini:2013mzu,Gando:2012zm,Elliott:2016ble,Andringa:2015tza,Agostini:2017iyd},
at the level  of $T_{1/2} >  5.3\times10^{25}$~y  for $^{76}$Ge~\cite{Agostini:2017iyd}
and  $T_{1/2} >  1.07\times10^{26}$~y  for $^{136}$Xe~\cite{KamLAND-Zen:2016pfg}, 
with next generation ton-scale experiments aiming at two orders of magnitude sensitivity improvements.

By itself, the observation of \NLDBD \   would not  immediately point to the underlying mechanism of  LNV. 
In an effective theory approach to new physics,  LNV arises from $\Delta L=2$ operators of odd dimension, 
starting at dimension-five~\cite{Weinberg:1979sa,Babu:2001ex,deGouvea:2007qla,Lehman:2014jma}.
As discussed in detail in Ref.~\cite{Cirigliano:2017djv}, 
if the scale of  lepton number violation, $\Lambda_{\rm LNV}$,  is in the range 1-100 TeV,  
short-distance  effects  encoded in local operators of  dimension seven and nine 
provide  contributions to \NLDBD\ within reach of next generation experiments.  
However, whenever  $\Lambda_{\rm LNV}$ is much higher than the electroweak scale,  
the only low-energy manifestation of this new physics is a Majorana mass 
for light neutrinos, encoded in a  single gauge-invariant dimension-5 operator~\cite{Weinberg:1979sa}, 
which induces \NLDBD \ through light Majorana-neutrino exchange~\cite{Bilenky:2014uka,Bilenky:1987ty,Haxton:1985am}.

To interpret positive or null \NLDBD \ results in the context of this minimal extension of the SM 
(the three light Majorana neutrinos paradigm),  it is  critical  to have good control over the 
relevant hadronic and nuclear matrix elements. Current knowledge of these is somewhat 
unsatisfactory~\cite{Engel:2016xgb}, as
(i) few  of the current calculations are  based on a modern effective field theory (EFT) analysis,  
and  
(ii) various approaches lead to estimates that differ by a factor of two to three.  
In this paper we present the first  end-to-end EFT analysis of \NLDBD\ induced by light Majorana-neutrino exchange, 
describing the physics from the scale $\Lambda_{\rm LNV}$ all the way down to the nuclear energy scale. 
The EFT framework has allowed us to identify long- and short-range contributions to \NLDBD\ previously missed in the literature, 
that are,  by power counting, as large as corrections usually included.  
The main results of our work are expressions for  
the leading and next-to-next-to-leading order (N$^2$LO)
chiral potentials mediating \NLDBD,  
and the amplitude induced by the  exchange of ultrasoft neutrinos, with momenta much smaller than the Fermi momentum.

\section{Effective theory framework}
The starting point of our analysis is
 the weak scale effective Lagrangian, which we take to be the SM  
augmented by Weinberg's  $\Delta L = 2$  dimension-five operator \cite{Weinberg:1979sa}, 
\be
{\cal L}_{\rm eff} =    {\cal L}_{\rm SM}   \ + \left\{\ \frac{u_{\alpha \beta}}{\Lambda_{\rm LNV}} \,   \epsilon_{ij} \epsilon_{mn}  L_i^{T \alpha} C L_m^\beta   \ H_j H_n +{\rm h.c.}\right\}\,\,,
\label{eq:Seff0.5}
\ee
where $u_{\alpha \beta}$ is a $3\times3$ matrix,
$L = (\nu_L \  e_L)^T$ is the left-handed $SU(2)$ lepton doublet,  $H$ is the Higgs doublet,  
$\alpha, \beta \in {e, \mu, \tau}$, and  $i,j,m,n$ are $SU(2)$ indices.  
This operator induces a Majorana mass  matrix for neutrinos, of the form $m_{\alpha \beta} =  -u_{\alpha \beta}  (v^2/\Lambda_{\rm LNV})$, 
where $v = (\sqrt{2} G_F)^{-1/2}  \simeq 246$~GeV is the Higgs vacuum expectation value:  for $\Lambda_{\rm LNV} \gg v$ this is the well known ``seesaw" relation. 
 
Neglecting QED and weak neutral-current effects, the low-energy 
effective Lagrangian at scale  $\mu   \gtrsim \Lambda_\chi \sim 1$~GeV  is given by
\be
{\cal L}_{\rm eff} =    {\cal L}_{\rm QCD}  \ -\left\{ \   2 \sqrt{2} G_F V_{ud} \    \bar{u}_L \gamma^\mu d_L \, \bar{e}_L \gamma_\mu \nu_{eL} 
\ + \   \frac{1}{2} m_{\beta \beta}  \  {\nu}_{eL}^T  C  \nu_{eL}   
\ - \     C_L    \,  O_{L} +{\rm h.c.}\right\}~. 
\label{eq:Seff0}
\ee
The second term in \eqref{eq:Seff0} 
represents the Fermi charged-current weak interaction. 
The last two terms encode LNV through the neutrino  Majorana  mass, 
given by $m_{\beta \beta} = \sum_i  U_{ei}^2  m_i$ in terms of mass eigenstates and elements of the neutrino mixing matrix, 
and  a dimension-nine $\Delta L = 2$ operator  
generated at the electroweak threshold: 
 $O_{L} =  \bar{e}_L     e_L^c \  \bar{u}_L \gamma_\mu d_L \  \bar{u}_L  \gamma^\mu d_L $, with $e_L^c = C \bar{e}_L^T$. 
 Since 
 $C_L =    (8 V_{ud}^2  G_F^2 m_{\beta \beta})/M_W^2   \times  (1 + \mathcal O(\alpha_s/\pi)) $, 
the effect of the latter term on the \NLDBD \ amplitude is suppressed by $(k_F/M_W)^2$  (where $k_F \sim\mathcal  O(100)$~MeV is the typical Fermi momentum of nucleons in a nucleus) 
compared to light-neutrino exchange and can be safely neglected at this stage.  

The interactions of Eq.~\eqref{eq:Seff0} induce  $\Delta L= 2$  transitions (such as  $\pi^- \pi^- \to e^- e^-$, $n n  \to  p p e^- e^-$, 
$^{76}{\rm Ge} \to  \,  ^{76}{\rm Se} \,  e^- e^-$,  $^{136}{\rm Xe} \to \,    ^{136}{\rm Ba} \, e^- e^-$, ....) through  the non-local effective action  
obtained by contracting the neutrino fields in the two weak vertices, 
\be
S_{\rm eff}^{\Delta L = 2} 
=  \frac{ 8  G_F^2 V_{ud}^2  m_{\beta \beta} }{2!}    \int d^4 x  d^4y    \ S(x-y) \times   \bar{e}_L   (x)  \gamma^\mu  \gamma^\nu e_L^c (y)  
\times  T \Big( \bar{u}_L \gamma_\mu d_L(x)   \ \bar{u}_L \gamma_\mu d_L(y)  \Big)\,\,,
\label{eq:Seff1-v0}
\ee
where
\be  \label{eq:EffProp}
S(r) = \int \frac{d^4q}{(2 \pi)^4}   \frac{e^{-i q \cdot r}}{q^2  + i \epsilon}
\ee
is the scalar massless propagator. 
Computing matrix elements of $S_{\rm eff}^{\Delta L=2}$ in hadronic and nuclear states
is a notoriously difficult task.  
The multi-scale nature of the problem can be seen 
more explicitly by going to the Fourier representation%
%FOOTNOTE
\footnote{To obtain \eqref{eq:Sfourier} we have approximated   
$\bar{e}_L   (x)  \gamma^\mu  \gamma^\nu e_L^c (y) \simeq  \bar{e}_L   (x)  \gamma^\mu  \gamma^\nu e_L^c (x)  = g^{\mu \nu}   \bar{e}_L   (x)  e_L^c (x)$, 
which amounts to neglecting  the difference in electron momenta,  a safe assumption given that $|p_1 - p_2|/k_F \ll 1$.} 
\bea
\langle e_1 e_2 h_f |  S_{\rm eff}^{\Delta L = 2}  |h_i \rangle 
&=&   \frac{ 8  G_F^2 V_{ud}^2  m_{\beta \beta} }{2!}    \int d^4 x  \  
\langle e_1 e_2 |    \bar{e}_L   (x)  e_L^c (x)  | 0 \rangle 
\int \frac{d^4k}{(2 \pi)^4} 
\frac{  g^{\mu \nu}  \hat{\Pi}_{\mu \nu}^{++} (k,x) }{ k^2  + i \epsilon}\,\,,
\label{eq:Sfourier}
\\
\hat{\Pi}_{\mu \nu}^{++} (k,x) &=&  \int d^4 r  \, e^{i k \cdot r} \  \langle h_f |  T  \Big(  \bar{u}_L \gamma_\mu d_L(x +r/2)   \ \bar{u}_L \gamma_\mu d_L(x - r/2)  \Big)
| h_i \rangle ~.
\label{eq:correlator}
\eea
The amplitude  \eqref{eq:Sfourier}  receives contributions from neutrino virtualities ranging from the weak scale all the way down to the 
IR scale of nuclear bound states.  
Roughly speaking one  can  identify  three regions,  
whose contributions can be conveniently described in terms of appropriate effective theories:   

(i) A hard region  with $k_E^2 \equiv  (k^0)^2  + \vec{k}^2  \gg  \Lambda_\chi^2  \sim  1 \,  {\rm GeV}^2$. 
This contribution is controlled by the  quark-level short-distance behavior of the correlator \eqref{eq:correlator}. 
An    Operator Product Expansion analysis shows that  integrating out hard neutrinos and gluons 
generates a local term in the effective action  proportional to $O_{L}$,  with Wilson coefficient 
\be\label{eq:4q}
C_{L}  (\Lambda_\chi) =   \frac{8 G_F^2 V_{ud}^2  m_{\beta \beta}}{\Lambda_\chi^2}  \, \frac{ \alpha_s (\Lambda_\chi)}{4 \pi}~.
\ee  
This short-distance component is currently missing in all calculations of \NLDBD,  which start from the nucleon-level 
realization of the weak currents in the  correlator \eqref{eq:correlator}.  
Within such approaches,  the new effect  can be  estimated by considering the hadronic realization of  $O_{L}$, 
sensitive to pion-range and short-range nuclear effects,  
that has been studied in the context of TeV-sources of LNV~\cite{Savage:1998yh,Prezeau:2003xn,
Nicholson:2016byl,Cirigliano:2017ymo}.  
In what follows we adopt a chiral EFT approach and the effect of  hard modes 
will be encoded in  local counterterms of the low-energy effective chiral Lagrangian, transforming as $O_{L}$ under the chiral group.~\footnote{Within Lattice QCD, $O_L$ 
captures  $\mathcal{O} (a^2)$  discretization effects 
in the calculation of the amplitude~\eqref{eq:Sfourier}.   
$O_L$  would appear in the Symanzik's action~\cite{Symanzik:1983dc,Symanzik:1983gh} 
with a pre-factor scaling as $\mathcal{O}(\alpha_s a^2)$  near the  continuum limit.   
Similar contributions relevant to the case of two-neutrino double beta decay  ($2\nu\beta\beta$)  have been discussed in Ref.~\cite{Tiburzi:2017iux}.}

(ii)  A soft and potential region with $k_E^2 \sim k_F^2  <  \Lambda_\chi^2$. 
Here the  appropriate hadronic degrees of freedom  are pions and nucleons, described by chiral EFT. 
In  analogy with the strong and electroweak interactions in the SM, 
integrating out pion degrees of freedom and neutrinos  with 
soft ($k^0 \sim |\vec{k}| \sim m_\pi \sim  k_F$)  and   potential  ($k^0 \sim  k_F^2/m_N$,  $|\vec{k}| \sim k_F$)   scaling of their 4-momenta
generates  nucleon level  $\Delta L =  \Delta I = 2$ potentials  that mediate \NLDBD \  between nuclear  states. 

(iii)  Ultrasoft or ``radiation"  region,  with neutrino momenta scaling as  $k_0\sim |\vec k|  \ll k_F$. 
Here the  effective theory contains as explicit degrees of freedom nucleons interacting 
via appropriate potentials (see (ii) above), electrons, and essentially massless neutrinos, 
whose ultrasoft modes cannot be integrated out (similarly to gauge fields in 
NRQED and NRQCD~\cite{Pineda:1997bj,Pineda:1997ie,Brambilla:1999xf}).
These modes do not resolve the nuclear constituents and this part of the amplitude is sensitive 
to nuclear  excited states and transitions among them induced by the electroweak currents. 

Contributions to \NLDBD \  from regions (ii) and (iii) are 
included  in all existing calculations albeit within certain approximations and not fully in the spirit of EFT. 
In particular, we have identified  corrections that can not be parameterized through the single nucleon form factors.
We next discuss the \NLDBD \ amplitude in the context of chiral EFT,  in which 
the contributions from region (i) are captured by local counterterms, 
the  contributions from region (ii)  can be explicitly evaluated and lead to appropriate potentials, 
and the contributions from region (iii) can be  displayed in terms of non-perturbative nuclear matrix 
elements of the weak charged current  and bound state energies.

\section{Chiral EFT and \NLDBD}

We describe the low-energy  realization of the GeV-scale effective Lagrangian in Eq.~\eqref{eq:Seff0} 
in the framework of chiral perturbation theory ($\chi$PT) \cite{Weinberg:1978kz,Gasser:1983yg,Jenkins:1990jv,Bernard:1995dp}  and its generalization to multi-nucleon systems, chiral EFT \cite{Weinberg:1990rz,Weinberg:1991um,Ordonez:1992xp}.

Chiral symmetry and its spontaneous and explicit breaking strongly constrain the form of the interactions among
nucleons and pions. In the limit of vanishing quark masses,  the $\chi$PT Lagrangian is  obtained by constructing all chiral-invariant interactions between nucleons and pions.
Pion interactions are derivative, allowing for an expansion in 
$p/\Lambda_\chi$, where $p$ is the typical momentum scale in a process and $\Lambda_\chi \sim m_N \sim 1$ GeV 
is the intrinsic mass scale of QCD. 
One can order interactions according to the chiral index $\Delta = d + n/2-2$, where $d$ counts the number of 
derivatives and $n$ counts the number of nucleon fields \cite{Weinberg:1978kz,Weinberg:1990rz}.
Chiral symmetry is explicitly broken by the quark masses and charges, and, in our case, by electroweak and $\Delta L=2$ operators.  
However, the explicit breaking is small and can be systematically included in the power counting by considering  $ m_q\sim\mpi^2\sim p^2$.
In presence of lepton fields we  generalize the definition of  chiral index to 
$\Delta = d + n/2-2 +n_e$,  where $n_e$ denotes the number of charged leptons in the interaction vertex. 
With this definition, the lowest order $0\nu\beta\beta$ transition operators have chiral index $\Delta = 0$.
For nuclear physics applications,  one has $p \sim k_F \sim m_\pi$  and the expansion parameter is  
$\epsilon_\chi = m_\pi/\Lambda_\chi$. 
For \NLDBD \ there are additional infrared scales.  
The energy  differences $E_n - E_i$  of the bound nuclear states 
have typical size  $\mathcal O(5-10)$ MeV,
to which we assign the scaling $k_F^2/m_N \sim k_F \,  \epsilon_\chi$.
For the reaction  $Q$ value and the electron energies $E_{1,2}$ the scaling $Q\sim E_{1,2}\ \sim k_F \,  \epsilon^2_\chi$,  was found to work well in Ref.\ \cite{Cirigliano:2017djv}.\

Our building blocks are the pion field   $u   = \exp\left(  i \boldpi   \cdot \boldtau / (2 F_0) \right)$ 
(where $F_0$ is the pion decay constant in the chiral limit, and  $F_\pi=92.2$ MeV)  
and the nucleon doublet $N= (p\, n)^T$,   transforming as  $u\to LuK^\dagger (\pi) =K (\pi) uR^\dagger$ and  $N\to K(\pi)N$  
under the  $SU(2)_L\times SU(2)_R$ chiral group~\cite{Coleman:1969sm,Callan:1969sn}. 
 The effective Lagrangian of Eq.~\eqref{eq:Seff0} maps onto the following operators 
with zero chiral index,  
\begin{subequations}
\label{eq:LchiralLO}
\bea
\vL_\pi^{(0)} &=& \frac{F_0\sq}{4}{\rm Tr}\left[  u_\mu u^\mu + u^\dagger \chi u^\dagger   + u \chi^\dagger u  \right] ~ ,
\qquad  u_\mu = - i \left[ u^\dagger   ( \partial_\mu  - i l_\mu) u - u  \partial_\mu   u^\dagger  \right] \,\,, 
\\
\mathcal L^{(0)}_{\pi N} &=&  i\bar N  v^\mu   (\partial_\mu + \Gamma_\mu )  N + g^0_A \bar N S^\mu  u_\mu  N ~,  
\qquad  \Gamma_\mu =  \frac{1}{2}  \left[ u^\dagger   ( \partial_\mu  - i l_\mu) u + u  \partial_\mu   u^\dagger  \right]  \,\,,
\\
\mathcal L^{(0)}_{N N} &=&   -  \frac{C_S}{2} \bar N  N \bar{N} N  -  \frac{C_T}{2} \bar{N}  \vec{\sigma} N \bar{N}   \vec{\sigma} N~, \label{eq:LchiralLOc}
\eea
\end{subequations}
where  $\chi = 2 B \times {\rm diag}(m_u,m_d)$ (with $B(\mu=2 \,{\rm GeV}) \simeq 2.8$~GeV), 
$ l_\mu =  - 2  \sqrt{2}  G_F V_{ud}    \tau^+    \   \bar e_L\g_\mu \nu_L +{\rm h.c.}$, 
$C_{S,T} = \mathcal O({F_0}^{-2})$,  
and $g^0_A$ is the LO contribution to the nucleon axial coupling which is measured to be 
$g_A=1.2723(23)$~\cite{Agashe:2014kda}.
Tree-level diagrams involving  the above interactions   and Majorana neutrino exchange  
generate $\Delta L = 2$  amplitudes such as $\pi^- \pi^- \to ee$ and  $nn \to pp ee$,
scaling as  $\mathcal O(G_F^2 m_{\beta \beta})$.
At the one-loop level UV divergences appear which require the introduction of  $\Delta L =  \Delta I = 2$ local operators 
with chiral index $\Delta = 2$. 
We find three independent structures with the correct transformation properties: 
\bea \label{eq:CT}
\vL^{(2)}_{\Delta L = 2} &=& \bigg\{ \frac{5}{12} F_0^4   g_\nu^{\pi \pi} \  L_{21}^\mu L_{21\,\mu} \,  
+  g_A^0 F_0^2    g_\nu^{\pi N} \  \bar N S^\mu u^\dagger\tau^+ u N\,{\rm Tr}\left(u_\mu u^\dagger \tau^+ u \right)   
\nn \\
&+&    g_\nu^{NN} \, (\bar N u^\dagger \tau^+ u N)(\bar N u^\dagger \tau^+ u N)   \bigg\}  \,  \kappa \,  \bar e_L C\bar e_L^T +{\rm h.c.}
\qquad \qquad \qquad \kappa = \frac{2 G_F^2 V_{ud}^2  m_{\beta \beta}}{(4 \pi F_0)^2}
\nn   \\
&=&
\left[    \frac{5}{6}  F_0^2   g_\nu^{\pi \pi} \,  \partial_\mu \pi^- \partial^\mu \pi^- 
+   \sqrt{2}g_A^0 F_0    g_\nu^{\pi N}  \   \bar p S_\mu n   \,  \partial^\mu  \pi^- 
+ g_\nu^{NN}   \bar p n \,  \bar{p} n   \right]  \  \kappa\,  \bar e_L C\bar e_L^T+\dots  \qquad 
\eea
Here $L^\mu =  u u^\mu u^\dagger $,  the dots stand for terms involving more than two pions, and 
three a priori unknown  $\mathcal O(1)$ low-energy constants (LECs) appear:   $g_\nu^{\pi \pi}$, $g_\nu^{\pi N}$,  and $g_\nu^{NN}$.

In the mesonic and single-nucleon sector of the theory, all momenta and energies are typically $\sim p$, and  the perturbative expansion of the $\chi$PT Lagrangian 
and power counting of loops~\cite{Weinberg:1978kz}  implies that the scattering amplitudes can also be expanded in $p/\Lambda_\chi$.
For systems with two or more nucleons
the energy scale  $p^2/2m_N$ becomes relevant 
and the corresponding amplitudes do not have a homogeneous scaling in $p$.  
Therefore, the perturbative expansion of  interactions   
does not guarantee a perturbative expansion of the amplitudes \cite{Weinberg:1990rz,Weinberg:1991um}.
Indeed, the  so-called ``reducible" diagrams (in which the intermediate state consists purely of propagating nucleons)  
are enhanced by factors of $m_N/p$ with respect to the $\chi$PT
power counting and need to be resummed. 
On the other hand, loop  diagrams whose intermediate states contain interacting nucleons and pions --``irreducible''-- 
follow the $\chi$PT power counting~\cite{Weinberg:1990rz,Weinberg:1991um}.
Reducible diagrams are then obtained by patching together irreducible diagrams with intermediate states consisting of $A$ free-nucleon propagators.
This is equivalent to solving the Schr\"{o}dinger equation with a potential $V$ defined by the sum of irreducible diagrams.
For external  perturbations, such as electroweak currents and $\Delta L=2$ interactions, one can similarly  
identify irreducible contributions, that admit an expansion in $p/\Lambda_\chi$ \cite{Bedaque:2002mn}. 

While the scaling of irreducible loop diagrams is unambiguous, the power counting for four-nucleon operators has been the object of much debate in the literature \cite{Kaplan:1996xu,Bedaque:2002mn,Nogga:2005hy}.    
In the Weinberg power counting~\cite{Weinberg:1990rz,Weinberg:1991um}, the scaling  is determined by naive dimensional analysis, and the lowest order four-nucleon  operators in the strong and $\Delta L=2$ sectors are given, respectively,
by $C_{S,T} \sim  \mathcal O(F_0^{-2})$ in Eq.~\eqref{eq:LchiralLOc} and $g_{\nu}^{NN} \sim \mathcal O(1)$ in Eq.~\eqref{eq:CT}.  
While phenomenologically successful \cite{Meissner:2015wva}, the Weinberg power counting is not fully consistent. Inconsistencies appear in some channels, such as the $^1S_0$ channel, where the cutoff dependence of the solution of the Lippman-Schwinger equation cannot be absorbed by the counterterms that appear at lowest order \cite{Kaplan:1996xu,Nogga:2005hy}.
Various solutions to this problem  have been proposed, including  treating pion exchange in pertubation theory (perturbative pion or ``KSW''  scheme \cite{Kaplan:1996xu,Kaplan:1998tg,Kaplan:1998we}),
expanding the nuclear forces around the chiral limit \cite{Beane:2001bc},  or, for processes at low enough energy, 
integrating out pions and working in the  pionless EFT \cite{Chen:1999tn,Bedaque:2002mn}.
In Section \ref{Weinberg}, we will discuss \NLDBD\ in the Weinberg power counting, and we will extend the treatment to the pionless EFT  in Section~\ref{sect:pionless}.

\begin{figure}
\begin{center}
\includegraphics[width=0.25\textwidth]{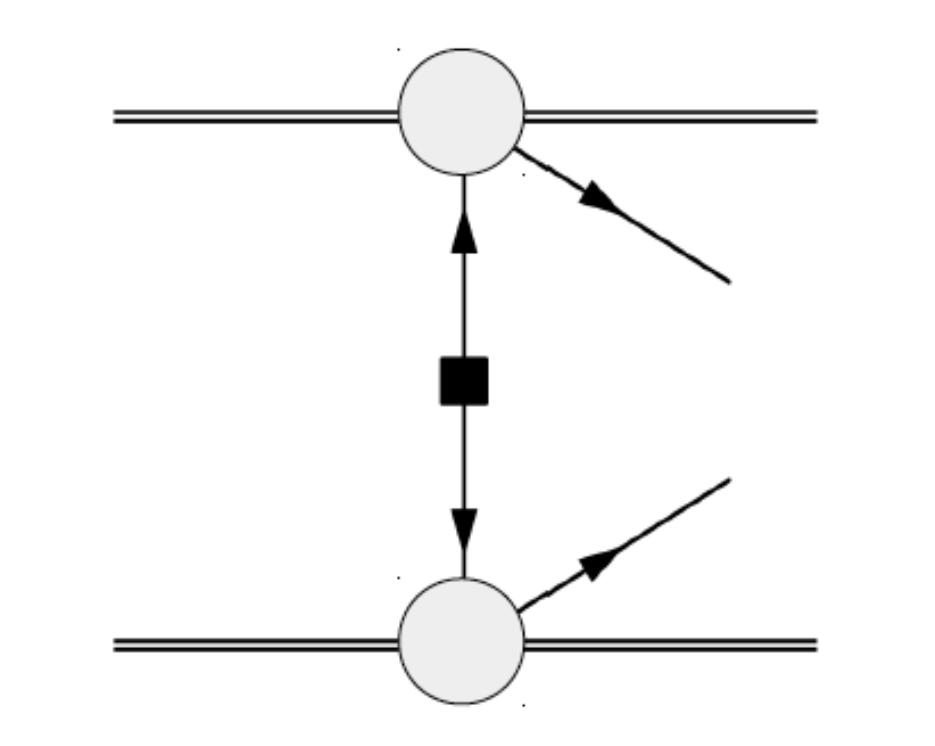}
\vspace{-0.75cm}
\end{center}
\caption{
 Diagram  contributing to the leading-order neutrino potential. Double and single lines denote, respectively, nucleons and lepton fields. The black square denotes an insertion of the neutrino 
Majorana mass, while the  gray circle denotes the SM weak charged-current interaction. 
}\label{twobody}
\end{figure}

Finally, matching the chiral EFT framework to many-body quantum mechanics  
one obtains the following  nuclear  Hamiltonian  appropriate for calculating  \NLDBD \ amplitudes:
\begin{equation}
H_{\rm eff} =   H_0  \ + \  \sqrt{2} G_F V_{ud}  \,
\sum_{n=1}^A   \left(   g_V \delta^{\mu 0} - g_A  \delta^{\mu i} \sigma^{(n)i}  \right)  \tau^{(n)+}   \   \bar{e}_L \gamma_\mu \nu_L   (x_n)
\ + \   2 G_F^2  V_{ud}^2 \  m_{\beta \beta}  \  \bar e_{L} C \bar e_{L}^T   \ V_\nu~.
\label{eq:Hpotential}
\end{equation}
The first term ($H_0$) encodes the strong interaction.   
In the Weinberg counting, the leading-order  strong potential is given by  one pion exchange plus the contact terms 
$C_{S,T}$~\cite{Weinberg:1990rz,Weinberg:1991um}, which in momentum space reads ($\vec{q}$ is the momentum conjugate to $ \vec{x}_{ab} 
\equiv  \vec{x}_a - \vec{x}_b$)
\be \label{strong}
V_{\rm strong, 0} = 
\frac{1}{2} 
\sum_{a\neq b} \ \left(- \frac{g_A^2}{4 F_\pi^2} \frac{\boldsigma^{(a)} \cdot \vec q \, \boldsigma^{(b)} \cdot \vec q }{\vec{q}^2 + m_\pi^2}   \boldtau^{(a)} \cdot \boldtau^{(b)}   \ 
\ + \ C_S  + C_T \,  \boldsigma^{(a)} \cdot  \boldsigma^{(b)}  \right) .
\ee
Here and in the following, we replace the LO couplings and decay constants by their physical values, $g^0_A\rightarrow g_A$, $F_0\rightarrow F_\pi$, etc., which can be consistently done to the order we are working in the chiral expansion.

The second term in \eqref{eq:Hpotential} 
is the usual charged-current weak interaction. 
From now on, we set $g_V=1$, neglecting small isospin-breaking corrections.
Note that light Majorana neutrinos and electrons with ultrasoft momenta 
are active degrees of freedom in the low-energy theory. 

The third term in \eqref{eq:Hpotential} directly mediates $\Delta L = 2$ amplitudes, and we discuss it next.

\subsection{The $\Delta L= \Delta I=2$ potential}\label{Weinberg}

The potential  $V_\nu$ encodes physics from hard scales (the counterterms  of Eq.~\eqref{eq:CT}) 
as well as soft scales,  obtained  by integrating out pions and Majorana neutrinos with 
soft and  potential  scaling of their 4-momenta.
In practice $V_\nu$ is given by the sum of 
 ``irreducible" diagrams mediating $nn \to pp ee$ in  chiral EFT.  
As discussed above,   $V_\nu$ admits  a chiral expansion:
\be
V_\nu =  \sum_{
a \neq b
}   \left(  V_{\nu,0}^{(a,b)}  + V_{\nu,2}^{(a,b)}  + \dots \right)~.
\label{eq:Vnu}
\ee

The LO neutrino potential is obtained by tree-level neutrino exchange,  which involves the single-nucleon currents 
(see Fig.~\ref{twobody}).   In momentum space
one finds~\cite{Cirigliano:2017djv}  
\begin{eqnarray}
    V_{\nu,0}^{(a,b)}  &=&    \tau^{(a)+} \tau^{(b)+}  
\,  \frac{1}{\vec{q}^2}  \,  \left\{
%g_V^2 -
1-   g_A^2  \left[
 \boldsigma^{(a)} \cdot \boldsigma^{(b)}    -  
  \boldsigma^{(a)} \cdot \vec q\,  \boldsigma^{(b)}\cdot \vec q
\    \frac{2 m_\pi^2  + \vec{q}^2}{(\vec q^2 + m_\pi^2)^2}
 \right]   \right\}~.
 \label{eq:Vnu0}
\end{eqnarray}
Analogously to the strong-interaction case, the 
neutrino potential $V_\nu$  depends only on the momentum scale $q \sim k_F$ 
and not  on infrared scales
such as the  excitation  energies of the intermediate 
odd-odd nucleus in \NLDBD,   
often approximated by their average  $\bar{E} - 1/2(E_i + E_f)$.
Note that the commonly used  neutrino potential~\cite{Bilenky:2014uka,Engel:2016xgb}   
reduces to $V_{\nu,0}$   
when $\bar{E} - 1/2(E_i + E_f)$ is set to zero.

\begin{figure}
\center
\includegraphics[width=0.75\textwidth]{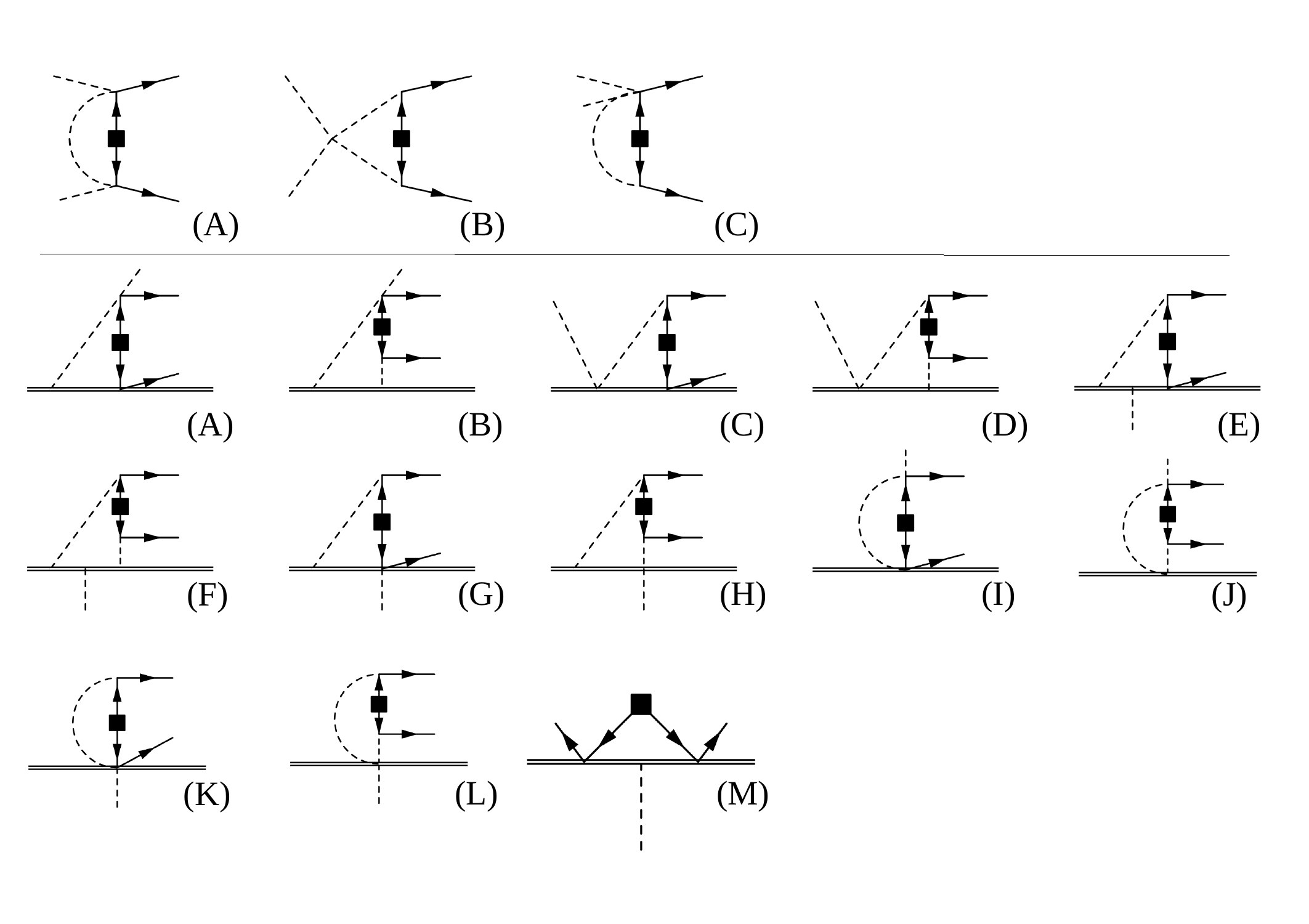}
\caption{Loop diagrams contributing to an effective $\pi \pi e^- e^-$ vertex (upper panel),
and to an effective $n p \pi e^- e^-$ vertex (lower panel). Pions are denoted by dashed lines, the remaining notation is as in Fig. \ref{twobody}. 
The diagrams give rise to corrections to the $\Delta L=2$ potential when the pions are connected to external nucleon lines. 
}\label{Fig1}
\end{figure}

\begin{figure}
\centering
\includegraphics[width=0.75\textwidth]{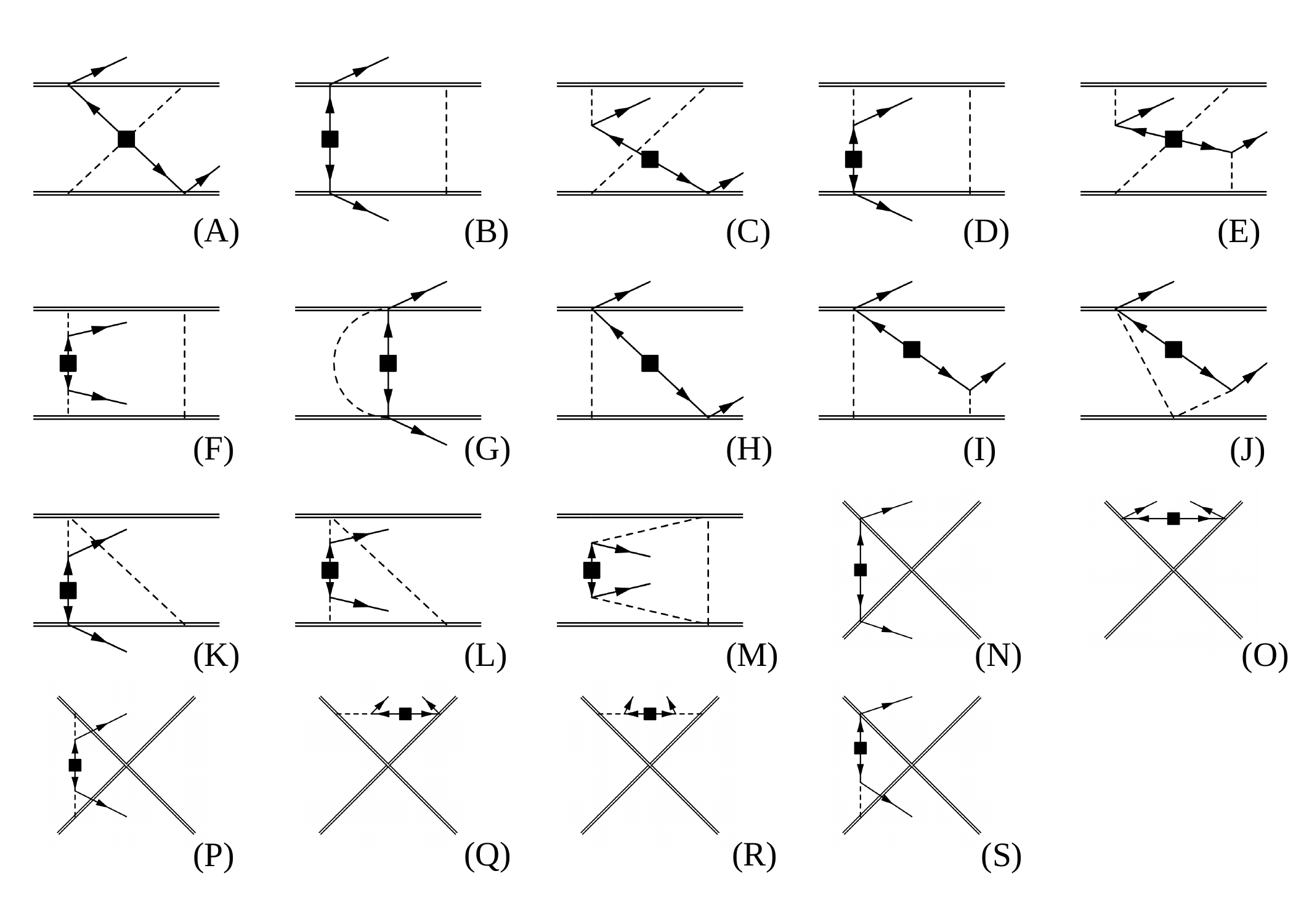}
\caption{Loop diagrams contributing to an effective $n p n p  e^- e^-$ vertex.}
\label{Fig2}
\end{figure}

\begin{figure}
\centering
\includegraphics[width=0.85\textwidth]{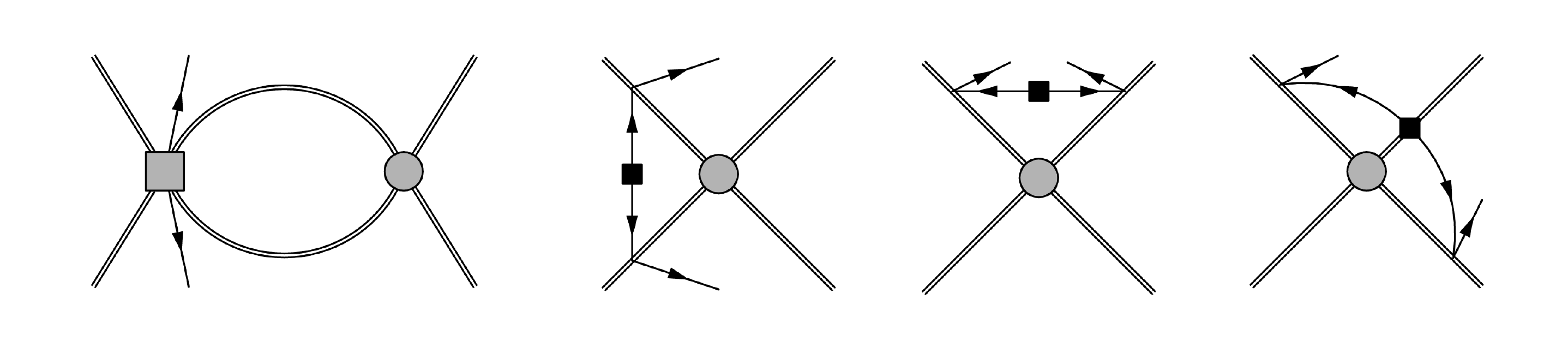}
\caption{
Diagrams in the low-energy nuclear EFT contributing to the matching at N$^2$LO.
The gray circle denotes an insertion of the LO  strong  potential of Eq.~\eqref{strong}. 
The gray box denotes an insertion of the LO  $\Delta L=2$ potential $V_{\nu, 0}$.
The remaining notation is as in Fig.~\ref{twobody}.}
\label{FigMatch}
\end{figure}

At N$^2$LO in the Weinberg power counting several  new contributions arise.  
These consist of  (a)  corrections to single-nucleon currents, which are often included in the literature via momentum-dependent form factors;
(b) genuine N$^2$LO     two-body effects, such as loop corrections to Fig.~\ref{twobody},  which induce  the short-range neutrino potential 
$V_{\nu,2}$
that  has never been considered in the literature.   
Note that  two-nucleon effects in the weak currents~\cite{Menendez:2011qq,Engel:2014pha}, 
which induce three-nucleon potentials in Eq.\ \eqref{eq:Vnu},
start contributing to \NLDBD\ at N$^3$LO, once one takes into account that $S\cdot v=0$ 
and $v \cdot q  \simeq \mathcal O(  {k_F^2} /{m_N})$.

Including N$^2$LO corrections to the single-nucleon currents, the  potential $V_{\nu, 0}^{(a,b)}$ is modified as  
\begin{eqnarray}
    V_{\nu, 0}^{(a,b)}  &=&    \tau^{(a)+} \tau^{(b)+}  
\,  \frac{1}{\vec{q}^2}  \,  g_A\sq\left\{
 h_F(\vec q^2)/g_A\sq  - \boldsigma^{(a)}\cdot \boldsigma^{(b)}  \, h_{GT}(\vec q^2)   
-  S^{(ab)}\,  h_T(\vec q^2)  \right\}~,
 \label{eq:Vnu0formfactors}
\end{eqnarray}
where we have introduced the tensor operator  $S^{(ab)}= -\left( 3\,\boldsigma^{(a)} \cdot   \vec q \, \boldsigma^{(b)} \cdot   \vec q -  \vec q^2 \boldsigma^{(a)}\cdot \boldsigma^{(b)} \right)/\vec q^2$.
The functions $h_F$, $h_{GT}$, and $h_T$ are expressed in terms of the isovector vector, axial, induced pseudoscalar, and magnetic nucleon form factors as \cite{Engel:2016xgb}
\begin{eqnarray}\label{eq:hK(q)}
h_F(\vec q^2) &=&   g_V^2(\vec q^2)\,,\nn\\ 
h_{GT}(\vec q^2) &=&   g_A^2(\vec q^2) + g_P(\vec q^2) \, g_A(\vec q^2) \frac{\vec q^2}{3 m_N} + g^2_P(\vec q^2) \frac{\vec q^4}{12 m_N^2} + g_M^2(\vec q^2) \frac{\vec q^2}{6  g_A\sq m_N^2},\nn\\
h_{T}(\vec q^2)  &=&  -  g_P(\vec q^2) \, g_A(\vec q^2) \frac{\vec q^2}{3 m_N} - g^2_P(\vec q^2) \frac{\vec q^4}{12 m_N^2} +  g_M^2(\vec q^2) \frac{\vec q^2}{12g_A\sq m_N^2}  \,.
\end{eqnarray}
In the literature, the dipole parameterization of the vector and axial form factors is often used
\begin{equation}\label{dipole}
g_V(\vec q^2) =  \left(1 + \frac{\vec q^2}{\Lambda_V^2}\right)^{-2}, \qquad  g_A(\vec q^2) =  \left(1 + \frac{\vec q^2}{\Lambda_A^2}\right)^{-2}~, 
\end{equation}
with vector and axial masses $\Lambda_V = 850$ MeV and $\Lambda_A = 1040$ MeV. 
The magnetic and induced pseudoscalar form factors are then assumed to be given by
\begin{equation}\label{assumption}
g_M(\vec q^2) = (1 + \kappa_1) g_V(\vec q^2), \qquad g_P(\vec q^2) = -\frac{2 m_N g_A(\vec q^2)}{\vec q^2 + m_\pi^2},
\end{equation}
where $\kappa_1 = 3.7$ is the nucleon isovector anomalous magnetic moment.
Expanding Eqs.\ \eqref{dipole} and \eqref{assumption} for small $|\vec q|$, 
one recovers the LO and, for $g_A(\vec q^2)$, the N$^{2}$LO $\chi$PT  expressions of the nucleon form factors.
In the case of $g_V$, $g_P$ and $g_M$, the N$^{2}$LO  $\chi$PT results,  given for example in Ref. \cite{Bernard:1998gv}, deviate from Eqs. \eqref{dipole} and \eqref{assumption}. 
However, any parameterization that satisfactorily describes  the observed nucleon form factors can be used in the neutrino potential \eqref{eq:Vnu0formfactors}.

The potential $V_{\nu, 2}$ is   induced  by  one-loop  diagrams with a virtual neutrino and pions  
contributing to $nn \to pp ee$,  built out of the leading interactions of Eqs.~\eqref{eq:LchiralLO}. 
They can be separated into  three classes,  involving   the $\pi \pi \to ee$   (Fig.~\ref{Fig1}, upper panel),   $n \to p \pi^+  e e$ (Fig.~\ref{Fig1}, lower panel),  
and  $nn \to ppee$ (Fig.~\ref{Fig2}) effective vertices. Note that for diagrams such as  Fig.~\ref{Fig2}$(A)$ or \ref{Fig2}$(D)$ we include only the two-nucleon irreducible component. 
We regulate the loops dimensionally, with scale $\mu$, and subtract the divergences according to the  $\overline{\rm MS}$  scheme. 
The UV divergences are  absorbed by the counterterms of Eq.~\eqref{eq:CT}, which  cancel the $\mu$ dependence 
of the loops and also provide finite contributions.

$V_\nu$ can be thought of as the matching coefficient between the chiral EFT and  the low-energy  
nuclear EFT described by Eq.~\eqref{eq:Hpotential},  containing  non-local potentials and ultrasoft neutrino modes.
The matching is achieved by subtracting the low-energy theory diagrams  depicted  in Fig.~\ref{FigMatch}, 
involving ultrasoft neutrino exchange and insertions of the  LO strong and $\Delta L=2$  potentials, 
from the chiral EFT diagrams of Figs.~\ref{Fig1} and \ref{Fig2}. 
Since the two EFTs have the same IR behavior,  the IR divergences,  
stemming  from diagrams $(M)$ of Fig.~\ref{Fig1} 
and $(A),(B)$ of Fig.~\ref{Fig2}, 
cancel in the matching.
We have  checked this by regulating the IR divergences with a neutrino mass.
Moreover,  ultrasoft  neutrino loops in Fig.~\ref{FigMatch}
contain UV divergences, 
which we  deal with in dimensional regularization and $\overline{\rm MS}$ subtraction, 
with renormalization scale $\mu_{\rm us}$.  Thus the matching leads to a term in the 
potential  $V_{\nu,2}^{(a,b)}$   that depends logarithmically on $\mu_{\rm us}$.  
As we show in Section~\ref{sect:amplitude} below, the dependence on $\mu_{\rm us}$ cancels once one includes the  ultrasoft contribution to the \NLDBD\ amplitude. 

Since we are interested in potentials  that mediate $0^+ \to 0^+$ nuclear transitions,  
we only need the parity-even contributions that arise from two insertions of the vector current (VV) or axial current (AA) and we write the 
N$^2$LO two-body potentials as:
\be
V_{\nu,2}^{(a,b)} =    \tau^{(a)+} \tau^{(b)+}  
 \, \left( {\cal V}_{VV}^{(a,b)}    + {\cal V}_{AA}^{(a,b)}   +  \tilde{\cal V}_{AA}^{(a,b)}   \, \log \frac{m_\pi^2}{\mu_{\rm us}^2}    
 + {\cal V}_{CT}^{(a,b)}    \right)  
\label{eq:Vnu2}.  
\ee
For the contribution of two vector currents, we find 
\be
{\cal V}_{VV}^{(a,b)}   =  -
\frac{g_A^2 }{(4\pi F_\pi)^2}    \frac{ \boldsigma^{(a)} \cdot \vec q \, \boldsigma^{(b)} \cdot \vec q }{m_\pi^2} 
\times  
 \left\{ \frac{2 (1 - \hat{q})^2}{\hat{q}^2 (1 + \hat{q})} \log\left( 1 + \hat{q} \right)  - \frac{2}{\hat{q}} + \frac{7 -  3 \hat{q}  L_\pi  }{(1 + \hat{q})^2}  + \frac{ L_\pi  }{1 + \hat{q}}\right\},
\label{eq:VV}
\ee
where $\hat{q} = - q^2/m_\pi^2$ and  $L_\pi = \log \frac{\mu^2}{m_\pi^2}$. 
 This form agrees with Ref.~\cite{vanKolck:1997fu}, where the virtual photon corrections to the one-pion exchange potential 
 were calculated.  For the axial component we find 
\bea
{\cal V}_{AA}^{(a,b)}  &=&   
\frac{g_A^2}{(4\pi F_\pi)^2} 
\frac{\boldsigma^{(a)} \cdot \vec q \, \boldsigma^{(b)} \cdot \vec q }{m_\pi^2} 
\left\{   \frac{g_A^2}{1 + \hat{q}} \left( L_\pi - 4  \right) +  \frac{1}{(1+\hat q)^2}  \right\} 
\label{eq:AA}
\\
& +&    \frac{\mathbf{1}^{(a)} \times \mathbf{1}^{(b)}}{(4\pi F_\pi)^2}  
\left\{   -\frac{3}{4} (1-g_A^2)^2 L_\pi    + g_A^4 f_4(\hat{q}) + g_A^2 f_2(\hat{q}) +  f_0(\hat{q}) +24g_A\sq F_\pi\sq C_T\left\{L_\pi+1\right\} \right\} \, 
\nonumber 
\\
\tilde{\cal V}_{AA}^{(a,b)} &=& 2 \,  \frac{g_A^4}{(4\pi F_\pi)^2}  \,  \frac{\boldsigma^{(a)} \cdot \vec q \, \boldsigma^{(b)} \cdot \vec q  \ + \
 \vec{q}^2 \  \mathbf{1}^{(a)}  \times \mathbf{1}^{(b)} }{\vec{q}^2 + m_\pi^2} -\frac{g_A^2}{(4\pi)^2}\, 48 C_T\, \mathbf{1}^{(a)}  \times \mathbf{1}^{(b)}\,\,,
\label{eq:Vtilde}
\end{eqnarray}
where 
\begin{eqnarray}
f_0(\hat q)  &=& -\frac{1 + 8 \hat{q}}{6 \hat{q}}  + \frac{(1 + \hat{q})(1 + 8 \hat{q} + \hat{q}^2 )}{6 \hat{q}^2} \log(1 + \hat{q}) - \frac{1}{24} (4 + \hat{q})(5 + 2 \qhat) g(\hat q) \\
f_2(\hat q)  &=&  \frac{1 + 8 \hat{q}}{3 \hat{q}}  + \frac{(1 + \hat{q})^2(-1 + 5 \hat{q})}{3 \hat{q}^2} \log(1 + \hat{q}) - \frac{1}{12} ( 40 + 47 \hat{q} + 10 \hat{q}^2  ) g(\hat q) \\
f_4(\hat q)  &=& - \frac{1}{6} \left(20 + \frac{1}{\qhat} - \frac{12}{4 + \qhat}\right)  - \frac{-1 + 14 \qhat + 78 \qhat^2 + 62 \qhat^3 + 23 \qhat^4}{6 \hat{q}^2 (1 + \qhat)} \log(1 + \hat{q})  \nonumber  \\
& & + \frac{1}{24 (4 + \hat{q})} (640 + 912 \qhat + 375 \qhat^2 + 46 \qhat^3) g(\hat q) , 
\end{eqnarray}
and the loop function $g(\hat q)$ is 
\begin{equation}
g(\hat q) = \frac{4}{\sqrt{\qhat (4 + \qhat) }} \textrm{arctanh} \left(\sqrt{\frac{\qhat}{4 + \qhat}}\right)~.
\end{equation}
Finally, the counterterm  potential reads
\be
{\cal V}_{CT}^{(a,b)}  
=\frac{g_A^2}{(4 \pi F_\pi)^2} \,   \frac{\boldsigma^{(a)} \cdot \vec q \, \boldsigma^{(b)} \cdot \vec q }{m_\pi^2} 
\left[ 
\frac{5}{6}  g_\nu^{\pi \pi} \, \frac{\hat{q}}{(1 + \hat{q})^2}    - g_\nu^{\pi N} \frac{1}{1+\hat q} 
\right] 
-   \frac{2 g_\nu^{NN}}{(4 \pi F_\pi)^2} \,     \mathbf{1}^{(a)} \times \mathbf{1}^{(b)}~. 
\ee
The $\mu$ dependence of   $g_\nu^{\pi \pi}$, $g_\nu^{\pi N}$,  and $g_\nu^{NN}$ cancels the $\mu$ dependence of $L_{\pi}$ in 
Eqs.~\eqref{eq:VV} and \eqref{eq:AA}. 
We  will discuss strategies to estimate the finite parts of the LECs in Section~\ref{sect:LECs}  below.

\subsection{The $\Delta L= \Delta I=2$ potential in the pionless EFT}
\label{sect:pionless}
The previous discussion assumed the Weinberg power counting, that, while phenomenologically successful \cite{Meissner:2015wva}, is not formally consistent \cite{Kaplan:1996xu,Nogga:2005hy}. 
Few-body systems and processes characterized by scales $p \ll m_\pi$ can be studied in pionless EFT, a low-energy EFT in which pion degrees of freedom are integrated out
\cite{Chen:1999tn,Bedaque:2002mn}. 
For physical pion masses, pionless EFT converges very well for the $A=2,3$ systems, 
and works satisfactorily well for up to $A=6$ \cite{Stetcu:2006ey}.    
While the application of this EFT to nuclei with $A>6$ needs to be studied in more detail,
it is interesting to extend the framework developed in the previous section to pionless EFT, 
especially in the light of a possible matching to lattice calculations of \NLDBD\, matrix elements performed at  heavy pion masses.
A similar matching between lattice and pionless EFT for strong interaction and  electroweak processes has been carried out in Refs.~\cite{Barnea:2013uqa,Kirscher:2015yda,Beane:2015yha,Savage:2016kon,Contessi:2017rww,Shanahan:2017bgi,Tiburzi:2017iux}.
While the lattice QCD calculations relevant for 2$\nu \beta \beta$ were performed at  a single lattice spacing of $a\sim0.145$~fm and at 
 $m_\pi\sim806$~MeV~\cite{Shanahan:2017bgi,Tiburzi:2017iux},  they represent the first step for the field and are very promising.  
We are optimistic that in the near future lattice calculations of electroweak processes and \NLDBD\ in the two nucleon system 
will  reach control over all lattice systematics,  as recently achieved for   the nucleon axial coupling  $g_A$~\cite{Bhattacharya:2016zcn,Bouchard:2016heu,Berkowitz:2017gql,callat:gA}.

Pionless EFT describes physics at the scale $p$ smaller than the cutoff of the theory $\Lambda_{\slashpi} \sim m_\pi$.
For power counting purposes, we introduce the scale $\aleph \sim p \ll \Lambda_{\slashpi}$.
The leading-order Lagrangian is given by Eq.\ \eqref{eq:LchiralLOc}, and the fine tuning of the $S$-wave nucleon-nucleon scattering
lengths is accounted for by assigning the coefficients $C_{S,T}$ the scaling  
\begin{equation}
C_{S,T}  =\mathcal O\left( \frac{4\pi}{m_N \aleph } \right).
\end{equation}
Using dimensional regularization with power divergence subtraction (PDS) \cite{Kaplan:1998we}, at the scale $\mu$ 
the couplings $C_{S,T}$ can be expressed in terms of the spin-singlet $^{1}S_0$ and spin-triplet $^3S_1$ scattering lengths $a_s$ and $a_t$ according to
\begin{equation}\label{scatt_len}
C_s = C_S -3 C_T  = \frac{4\pi}{m_N (a_s^{-1} - \mu)}, \qquad C_t = C_{T} + C_S =  \frac{4\pi}{m_N (a_t^{-1} - \mu)},
\end{equation}
where $a_s^{-1} \sim - 8.3$ MeV, $a_t^{-1} = 36$ MeV.  
Higher-order operators involve additional derivatives  and are 
related to additional parameters (effective range, shape parameter, \ldots) of the effective-range expansion.
Note that in pionless EFT the three-body nucleon force is a leading-order effect \cite{Bedaque:2002mn}.

\begin{figure}
\includegraphics[width=8cm]{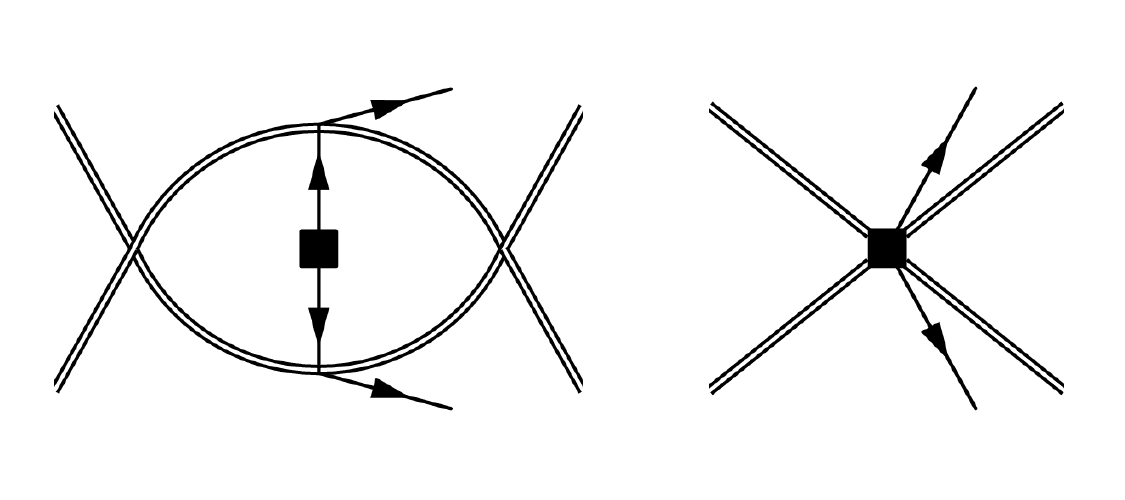}
\caption{Contributions to the $\Delta L =2$  $n n \rightarrow p p e^- e^-$ scattering amplitude in the pionless EFT. 
At leading order, the two neutrons (protons) in the initial (final) state have to be dressed by insertions of   $C_s$.}\label{piless}
\end{figure}

The leading $\Delta L=2$ potential in the pionless EFT has the form 
\begin{eqnarray}
    V_{\nu,0}^{(a,b)}  &=&    \tau^{(a)+} \tau^{(b)+}  
\,   \,  \left\{
\frac{1}{\vec{q}^2}  \left( g_V^2 -  g_A^2  
 \boldsigma^{(a)} \cdot \boldsigma^{(b)}   \right) -   \frac{2 g_\nu^{NN}}{(4 \pi F_0)^2} \,     \mathbf{1}^{(a)} \times \mathbf{1}^{(b)} \right\} .
 \label{eq:Vnu0piless}
\end{eqnarray}
The first term comes from long-range neutrino exchange,
as in Eq.\ \eqref{eq:Vnu0}, with the difference that the contributions of the induced pseudoscalar form factor are subleading. 
In addition, the scaling of the nucleon-nucleon  coupling $g_{\nu}^{NN}$, 
introduced in Eq.~\eqref{eq:CT},
is modified. This coupling connects two $S$-waves  and thus is enhanced in the pionless theory
\cite{Bedaque:2002mn}, scaling as
\begin{equation}\label{scalingpiless}
g_{\nu}^{NN} = \mathcal O\left(\frac{\Lambda_\chi^2}{\aleph^2}\right).
\end{equation}
This scaling can be understood by studying the scattering amplitude for two neutrons to turn into two protons  with the emission of two zero-momentum electrons. 
At leading order in the pionless EFT, the scattering amplitude in the $^1S_0$ channel receives contributions from the diagrams  in Fig.~\ref{piless}, 
where the two neutrons and two protons in the initial and final states are dressed by bubble diagrams with insertions of the leading order contact interaction $C_s$.
These contributions to the amplitude  have the schematic form
\begin{equation}\label{amp}
\mathcal A\left( nn (^{1}S_0) \rightarrow pp (^1S_0)\right)  \sim G_F^2 m_{\beta\beta} \left\{  \left(\frac{T}{C_s(\mu)}\right)^2 \left( \left(\frac{m_N C_s(\mu)}{4\pi} \right)^2 (1 + 3 g_A^2)\, I_2 - \frac{2 g_\nu^{NN}}{(4\pi F_0)^2}    \right) + \ldots
\right\}, 
\end{equation}
where $T$ is the leading-order, strong-interaction scattering amplitude in the $^{1}S_0$ channel, which is scale independent, and 
the dots in Eq.\ \eqref{amp} denote additional scale-independent contributions.
$I_2$ is the dimensionless two-loop integral that appears in the first diagram of Fig.\ \ref{piless}.
The loop is logarithmically divergent in $d=4$, giving, in the PDS scheme,
\begin{equation}
I_2 =  \frac{1}{2} \log \frac{\mu^2}{16\gamma^2} + \frac{1}{2}, \quad \gamma^2 = - m_N \left(E - \frac{\vec P^2}{4 m_N}\right),
\end{equation}
where $E$ is the energy of the two neutrons in the initial state, and $\vec P$ the center-of-mass momentum. 
This is the same UV divergence that appears in Coulomb corrections to proton-proton scattering \cite{Kong:1999sf}.
The amplitude \eqref{amp} can be made independent of the renormalization scale $\mu$ by rescaling 
\begin{equation}\label{rescaling}
g_\nu^{NN}(\mu) = (4\pi F_0)^2 \left(\frac{m_N C_s(\mu)}{4\pi}\right)^2 \tilde{g}_\nu^{NN}(\mu),
\end{equation}
where  $\tilde{g}_\nu^{NN} = \mathcal O(1)$   and satisfies
\begin{equation}
\frac{d}{d\log\mu}  \tilde{g}_\nu^{NN}  = \frac{1 + 3g_A^2}{2}.
\end{equation}%
Eq. \eqref{rescaling}, together with \eqref{scatt_len}, confirms the  power-counting expectation of Eq. \eqref{scalingpiless}.

Beyond leading order in the pionless EFT there appears a single  four-nucleon operator
contributing to $V_{\nu,2}$ at   N$^2$LO,  which is  conveniently  expressed in terms of the $^1S_0$ projectors $P_\pm^{(^1S_0)}$, 
\begin{equation}\label{piN2LO}
\mathcal  L =  g_{\nu,\, 2}^{NN} \kappa\,  \bar e_L C\bar e_L^T  \left\{ (N^T  \overleftrightarrow{\nabla}^{\,2} P^{(^1S_0)}_+   N) (N^T  P^{(^1S_0)}_-   N)^{\dagger} 
+ (N^T   P^{(^1S_0)}_+   N) (N^T \overleftrightarrow{\nabla}^{\,2} P^{(^1S_0)}_-   N)^{\dagger} 
\right\} 
+ \textrm{h.c.},   
\end{equation}
with $P^{(^1S_0)}_{\pm} = {(i \sigma_2) (i \tau_2\tau_{\pm})}/{2\sqrt{2}}$, and  $\kappa$ given in Eq.\ \eqref{eq:CT}. The LEC $g_{\nu,\, 2}^{NN}$ scales as  
\begin{equation}
g_{\nu,\, 2}^{NN} = \mathcal O\left(  \frac{\Lambda_\chi^2}{\aleph^2 \Lambda^2_{\slashpi}} \right).
\end{equation}
Operators connecting two neutrons and two protons in the $P$-waves are not enhanced by $\aleph^{-2}$, and appear at even higher order.

Additional corrections to the  potential $V_{\nu,0}$ arise from loop diagrams $(\ref{Fig2}N)$ and $(\ref{Fig2}O)$.
These diagrams are scaleless and vanish in dimensional regularization. If the infrared divergence is regulated by a neutrino mass $m_\nu$, the $m_\nu$ dependence is canceled
by the diagrams in Fig. \ref{FigMatch}, and one obtains
\be
V_{\nu, {\rm loops}}^{(a,b)} =     - \tau^{(a)+} \tau^{(b)+} \,    \left( \mathbf{1}^{(a)} \times \mathbf{1}^{(b)} \right) \, 
  \frac{48 g_A^2 C_T}{(4\pi)^2}    \, \log \frac{\mu^2}{\mu_{\rm us}^2}    ~
\label{eq:Vnu2piless}.  
\ee
The  $\mu$  dependence in~\eqref{eq:Vnu2piless} is reabsorbed by a sub-leading term in  $g_\nu^{NN}$.
Eq.~\eqref{eq:Vnu2piless} shows that  it is always possible to choose  $\mu_{\rm us}$ so that the correction to the potential 
vanishes, and the effect of diagrams  $(\ref{Fig2}n)$ and $(\ref{Fig2}o)$ is all encoded in the ultrasoft contribution.
Note that the loop~\eqref{eq:Vnu2piless}
and the  ultrasoft amplitude 
are suppressed by
$p/(4\pi m_N)   \sim (p/\Lambda_{\slashpi}) \times 1/(4 \pi)^2$ with respect to the  LO, and are thus smaller than corrections from Eq.~\eqref{piN2LO}, 
scaling as $p^2/\Lambda_{\slashpi}^2$.

The relevance of the ultrasoft region can be also understood diagrammatically. Indeed, while diagram $(\ref{Fig2}O)$ is suppressed with respect to the LO, diagrams with an arbitrary number of insertions of $C_s$ and
$C_t$ between the emission and absorption of the neutrino are not suppressed with respect to $(\ref{Fig2}O)$. These diagrams need to be resummed, and correspond to building up the intermediate states. They are captured by the ultrasoft contribution discussed in Section~\ref{sect:amplitude} below.

\subsection{The \NLDBD  \ amplitude}
\label{sect:amplitude}

\begin{figure}
\centering
\includegraphics[width=0.5\textwidth]{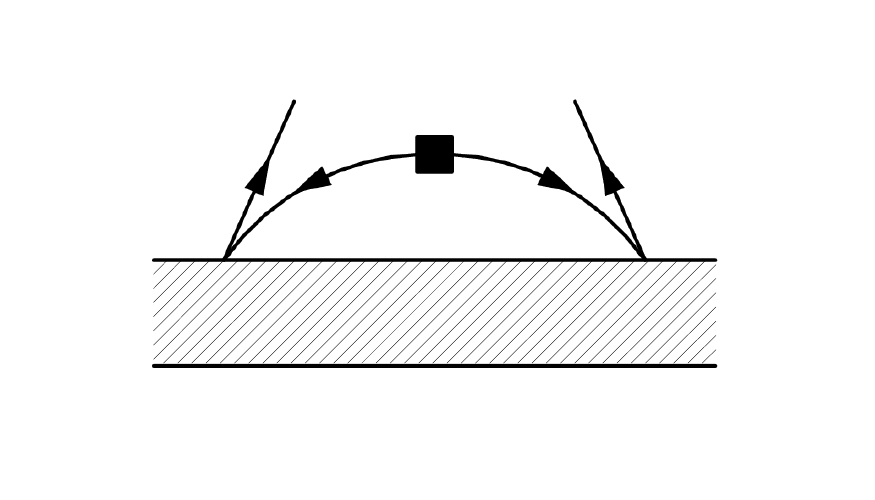}
\vspace{-0.75cm}
\caption{The ultrasoft contribution to \NLDBD\ amplitude. The  thick shaded  lines represent  nuclear bound states.
The remaining notation is as in Fig. \ref{twobody}.}
\label{FigBoundStates}
\end{figure}

Starting from the nuclear Hamiltonian of Eq.~\eqref{eq:Hpotential},  
one  calculates the full  \NLDBD\ amplitude as the sum of two contributions~\footnote{The amplitude $T_{fi}$ is related to the S-matrix element by 
$S_{fi} =  i (2 \pi)^4 \,  \delta^{(4)} (p_f - p_i)   \,  T_{fi}$.   Moreover   $(V_\nu)_{fi}$ is defined by 
$ \langle f |    V_\nu  |  i \rangle    =  (2 \pi)^3  \delta^{(3)}  \left(\vec{p}_f - \vec{p}_i \right)  \, \times (V_\nu)_{fi} $, 
where we have pulled out the 3-momentum delta function arising from integration 
over   the center-of-mass variables  that describe the overall motion of  $|i \rangle$ and $| f \rangle$.
$(V_\nu)_{fi}$ is related to the standard matrix element used in the \NLDBD\ literature~\cite{Engel:2016xgb} 
by   $(V_\nu)_{fi}  = -  (g_A^2/(4 \pi R_A)) (M_{GT} + M_T -  M_F/g_A^2)$, with 
$R_A= 1.2 A^{1/3}$~fm.}
\be
T_{fi} =  -  T_{\rm lept} \times      \Big( (V_{\nu,0})_{fi}    +  ( V_{\nu,2})_{fi}    \Big)  + T_{\rm usoft} ~, 
\label{eq:Tnldbd}
\ee
where we defined 
$T_{\rm lept} =  4  G_F^2 V_{ud}^2  m_{\beta \beta}  \bar{u}_L (p_{e1}) C \bar{u}_L^T (p_{e2})$.
The first term represents a single insertion of the $\Delta L = \Delta I = 2$ potential 
(third term in Eq.~\eqref{eq:Hpotential}).  
On the other hand   $T_{\rm usoft}$    arises from double insertions of the weak interaction 
(second term in Eq.~\eqref{eq:Hpotential}), which involves the exchange of ultrasoft Majorana neutrinos, with 
four-momenta scaling as $k^0 \sim |\vec{k}| \ll k_F$.   
The  diagram contributing to $T_{\rm usoft}$  is  given in Fig.~\ref{FigBoundStates} and its expression is 
\be
T_{\rm usoft} =  -\frac{T_{\rm lept}}{4}  \  \sum_n \   \int  \frac{d^{d-1}k}{(2 \pi)^{d-1}} 
\frac{1}{|\vec{k}|}  \, \left[
\frac{\langle f | J_\mu | n\rangle \langle n | J^\mu | i \rangle  }{|\vec{k}|   + E_2 + E_n - E_i - i \eta}  + 
\frac{\langle f | J_\mu | n\rangle \langle n | J^\mu | i \rangle  }{|\vec{k}|   + E_1 + E_n - E_i - i \eta}  
\right]~.
\label{eq:Tusoft1}
\ee
Here   $J_\mu \equiv V_\mu (\vec{x} =0)  - A_\mu (\vec{x}=0)
=  \sum_i  \, \tau^{(i)+}  \,  (\delta_{\mu 0} - g_A  \delta_{\mu k} \sigma^{(i)}_k) \, \delta^{(3)}  (\vec{x}_i)$
is the lowest-order nuclear weak current and 
$| n \rangle $ represent a complete set of nuclear states (eigenstates of $H_0$) 
with three-momentum $\pm \vec{k} + (1/2) (\vec{p}_i + \vec{p}_f)$ (the $\pm$ refer 
to the first and second term in  Eq.~\eqref{eq:Tusoft1}, respectively).   
The quantum numbers   of $J_\mu$ imply that, 
for given $0^+$ even-even initial and final states, 
$|n \rangle $ spans the set of 
eigenstates of the intermediate odd-odd nucleus.
Since we are in the ultrasoft regime, we  expand $\langle f | J_\mu | n\rangle \langle n | J^\mu | i \rangle$   in $\vec{k}$ 
and keep only the $\vec{k} = 0$ term,   noting  that  finite momentum terms would produce upon integration additional 
positive powers of the IR scale  $E_{1,2} + E_n - E_i$,  and therefore additional suppression.

Evaluating the loop integral in dimensional regularization with $\overline{\rm MS}$ subtraction, we find
\bea
T_{\rm usoft} (\mu_{\rm us}) &=&  T_{\rm lept}   \times \frac{1}{8 \pi^2}   \ 
\sum_n    \langle f | J_\mu | n\rangle \langle n | J^\mu | i \rangle  
\   \bigg\{  
  (E_2   + E_n  - E_i)     \left(  \log \frac{\mu_{\rm us}}{ 2( E_2 + E_n - E_i)}    + 1 \right)
\nonumber \\
&+&
 (E_1   + E_n  - E_i)     \left(  \log \frac{\mu_{\rm us}}{ 2( E_1 + E_n - E_i)}    + 1 \right)
\bigg\} ~,
\label{eq:Tusoft}
\eea
The UV divergence and the associated logarithmic dependence on $\mu_{\rm us}$ are  reabsorbed by 
the  term in the potential proportional to $\tilde{\cal V}_{AA}^{(a,b)}$.  
To verify  this,  using the completeness relation for the eigenstates of $H_0$ 
we write the term proportional to $\log \mu_{\rm us}$ in  \eqref{eq:Tusoft}  as a double commutator~\cite{Pineda:1997ie}
and evaluate it using  
the lowest order  chiral potential in $H_0$, finding
\begin{subequations}
\bea
\frac{d T_{\rm usoft}}{d \log \mu_{\rm us}}  
&=&  -  T _{\rm lept} \times \frac{1}{8 \pi^2}  \   \langle f |  \left[ J_\mu  ,  \left[ J^\mu,   H_0 \right] \right] | i \rangle 
=     T _{\rm lept} \times \frac{1}{8 \pi^2}  \   \langle f |  \left[ \vec{A}  ,  \left[  \vec{A},   H_0 \right] \right]  | i \rangle
\nonumber \\
&=&  -   2 \,  T _{\rm lept} \times   \sum_{a,b}    \langle f | \tau^{(a)+} \tau^{(b)+}   \  \tilde{\cal V}_{AA}^{(a,b)}  \ | i \rangle~,  
\\
- T_{\rm lept} \frac{d    (V_{\nu,2})_{fi}}{   d \log \mu_{\rm us}}
&=& 
+  2 T_{\rm lept} \times 
     \sum_{a,b}  \,  \langle f | \tau^{(a)+} \tau^{(b)+}   \  \tilde{\cal V}_{AA}^{(a,b)}  \ | i \rangle~,  
\eea
\label{eq:muus}%
\end{subequations}
with $\tilde{\cal V}_{AA}^{(a,b)}$ given in~\eqref{eq:Vtilde}.
The $\mu_{\rm us}$-independence of the total amplitude
implied by Eqs.~\eqref{eq:muus}  is a non--trivial consistency check for our calculation and 
allows us to pick a convenient scale, such as $\mu_{\rm us} = m_\pi$, which eliminates  the contribution of  $\tilde{\cal V}_{AA}$. 
Moreover,  the cancellation  implies  that $T_{\rm usoft}$ has the same chiral scaling as 
$(V_{\nu,2}^{(a,b)})_{fi}$, and is thus  two orders down  compared  to the leading contribution   
$(V_{\nu,0}^{(a,b)})_{fi}$.
This suppression can also be seen by  directly  comparing  the scaling of  $T_{\rm usoft}$ and $(V_{\nu, 0})_{fi}$.
In fact,   $(V_{\nu, 0})_{fi} \sim 1/(4 \pi R_{A} ) \sim k_F/(4 \pi)$~\footnote{
This   follows by taking matrix elements of  $V_{\nu,0} (r_{ab})  \propto  1/(4 \pi  r_{ab})$ 
between nuclear  states normalized to unity.}, 
which leads to 
$T_{\rm usoft} / T_0  \sim  \sum_n    (E_{1,2}  + E_n - E_i)/ (4 \pi k_F)    \times \langle f | J_\mu | n\rangle \langle n | J^\mu | i \rangle$.
Note that the  dimensionless transition matrix elements  $\langle f | J_\mu | n\rangle \langle n | J^\mu | i \rangle$  
also control the $2 \nu \beta \beta$ decay amplitude,   through a  different $E_n$-dependent  weighted sum~\cite{Engel:2016xgb}.
The overlap matrix elements quickly die out for $E_n - E_i > 10$~MeV, as borne out in several explicit calculations using 
different many-body methods~\cite{Coraggio:2017bqn,Caurier:2011gi,Simkovic:2010zw}.  
Therefore, recalling  the scaling  $E_n - E_i \sim k_F^2/m_N$ we recover 
$T_{\rm usoft} / T_0  \sim \epsilon_\chi^2$.

In summary,  in the chiral EFT framework one expects the following hierarchy 
of contributions to  the  \NLDBD \ amplitude of Eq.~\eqref{eq:Tnldbd}: 

\begin{itemize}

\item The leading contribution is given by   $T_0 =  -  T_{\rm lept} \times  (V_{\nu,0})_{fi} $
with  $V_{\nu,0}$ given in Eqs.~\eqref{eq:Vnu} and \eqref{eq:Vnu0}.  
This leading term is not sensitive to the intermediate states of the odd-odd nucleus. 
$V_{\nu,0}$ corresponds to the standard neutrino potential~\cite{Engel:2016xgb}  evaluated at  $\bar{E} - 1/2 (E_i + E_f)  \to  0$.

\item   A commonly included (but incomplete)  N$^2$LO contribution is  obtained  by inserting 
momentum dependent form factors in \eqref{eq:Vnu0}, 
as shown in  Eq.~\eqref{eq:Vnu0formfactors} and the subsequent discussion.

\item The new N$^2$LO contribution is given by  
$T_{2} =  -  T_{\rm lept} \times       ( V_{\nu,2})_{fi} (\mu_{\rm us} = m_\pi)  + T_{\rm usoft}$, 
with  $V_{\nu,2}$ given in Eqs.~\eqref{eq:Vnu}, \eqref{eq:Vnu2}   
and $T_{\rm usoft} (\mu_{\rm us})$ from Eq.~\eqref{eq:Tusoft}. 
With  the  choice  of  renormalization scale $\mu_{\rm us} = m_\pi$, $\tilde{\cal V}_{AA}$  drops out of the calculation.
Note that $T_{\rm usoft}$ requires the same nuclear structure input  needed in $2\nu \beta \beta$ calculations, namely  
$\langle f | J_\mu | n\rangle \langle n | J^\mu | i \rangle$  and the excited energy levels  of the intermediate nucleus ($E_n$).

\item In the pionless EFT one should use   $(V_{\nu,0})_{fi}$ from \eqref{eq:Vnu0piless} and 
$(V_{\nu,2})_{fi}$ from Eq.~\eqref{piN2LO}.  
Further suppressed contributions arise  from  \eqref{eq:Vnu2piless}  and 
$T_{\rm usoft} (\mu_{\rm us})$  in  Eq.~\eqref{eq:Tusoft}. 
Note that     $(V_{\nu, {\rm loops}})_{fi}$   in Eq.~\eqref{eq:Vnu2piless}  drops out 
when choosing  $\mu_{\rm us} = \mu$ with $\mu \sim \mathcal   O(m_\pi)$. 
 
 \end{itemize}

We suggest that many-body calculations be organized according to this  hierarchy, 
with the aim of (i) comparing results of  various methods order by order in chiral EFT 
and (ii)  checking to what degree  the chiral counting is respected in large nuclei. 

Finally, note that in evaluating $(V_{\nu,2})_{fi}$ in chiral EFT and pionless EFT  (and $(V_{\nu,0})_{fi}$ in pionless EFT), 
one encounters a priori unknown counterterms, 
which can be estimated in  naive dimensional analysis.
In the next section we discuss how to go beyond this rough estimate.

\subsection{Estimating the Low Energy Constants}
\label{sect:LECs}

\noindent {\bf Chiral EFT:} 
Interestingly,   $\pi \pi$ and $\pi N$ interactions similar to those in Eq.\ \eqref{eq:CT} are encountered when considering electromagnetic corrections to  
meson-meson and meson-nucleon interactions 
\cite{Urech:1994hd,Moussallam:1997xx,Gasser:2002am, Haefeli:2007ey,Ananthanarayan:2004qk}.
In  the electromagnetic  case, these operators arise from two insertions of the electromagnetic 
interaction, which involves the exchange of hard  photons.
In the case considered here,  the operators are generated by the exchange of hard neutrinos. 
However, the neutrino propagator 
and weak vertices  combine to give, up to a factor,  a massless  gauge boson  propagator  in Feynman gauge  (see Eqs.~\eqref{eq:Seff1-v0} and \eqref{eq:Sfourier}). 
This formal analogy can be exploited to relate the LECs needed for \NLDBD \
 (two insertions of the $\tau^+$ weak current)  
 to the  LECs    associated with the  $\Delta I = 2$ component of  the product of two electromagnetic currents,  that belongs to 
the  $5_L\times 1_R$ irreducible representation of the chiral SU(2) group.
Based on this observation, we  have identified  the operators of Refs.~\cite{Urech:1994hd,Moussallam:1997xx,Gasser:2002am, Haefeli:2007ey,Ananthanarayan:2004qk}  
that correspond to $g_\nu^{\pi\pi}$ and $g_\nu^{\pi N}$.
Explicitly, the relation between our couplings  renormalized  in the $\overline{\rm MS}$ scheme  and those of e.g.\ Ref.\ \cite{Gasser:2002am} 
(which are in the modified $\overline{\rm MS}$ scheme commonly employed in $\chi$PT 
\cite{Gasser:1984gg}), is given by
\bea
g_\nu^{\pi\pi} =- \frac{48}{5}(4\pi)^2
 \left(  \kappa_3^r  +    \frac{3}{8 (4 \pi)^2} \right)
\,\, ,\qquad g_\nu^{\pi N} = (4\pi F_\pi)^2 
\left(
\frac{g_4^r+g_5^r}{g_A}     -   \frac{1 - g_A^2}{(4 \pi  F)^2}
\right)
\,\,.
\eea
Our results for the anomalous dimensions of these couplings are in agreement with Ref.\ \cite{Urech:1994hd},
\bea
\frac{d g_\nu^{\pi\pi}}{d\ln \mu} =  - \frac{36}{5} \,\,, \qquad \frac{d g_\nu^{\pi N}}{d\ln \mu} = -2(1-g_A^2)\,\,.
\eea
At present the LEC $g_\nu^{\pi N}$ remains undetermined, while several estimates exist for $g_\nu^{\pi \pi}$~\cite{Baur:1996ya,Ananthanarayan:2004qk}. 
For example, Ref.~\cite{Ananthanarayan:2004qk}	finds  in Feynman gauge
 $\kappa_3^r  (\mu = m_\rho) = 2.7\cdot 10^{-3}$, which  
corresponds to $ g_\nu^{\pi \pi}   (\mu = m_\rho)= - 7.6$. We expect this estimate to be 
accurate at the 30-50\% level, as it relies on a large-$N_C$ inspired resonance  saturation of the correlators.
Finally, electromagnetic counterterms in the  two nucleon sector have been classified in Ref.~\cite{Walzl:2000cx}, 
but as far as we know no estimate of the finite parts exist, which would give us a handle on $g_\nu^{NN}$.

A first-principles  evaluation of  $g_\nu^{\pi \pi}$, $g_\nu^{\pi N}$,  $g_\nu^{NN}$  
based on Lattice QCD  is also possible. 
$g_\nu^{\pi \pi}$ and $g_\nu^{\pi N}$ can be determined by computing the  S-matrix elements  
for the processes   $\pi^- \pi^- \to ee$    and $n \to p \pi^+  e e$ on the lattice and matching to the corresponding 
chiral EFT expressions. 
On the lattice side,  one needs to compute matrix elements of the non-local effective action in Eq.~\eqref{eq:Seff1-v0} 
between appropriate external states.  As discussed above, the calculation is formally very similar
 (modulo the Lorentz and isospin structure of the currents)  to the 
one required to compute virtual photon corrections to hadronic processes.  Techniques being developed 
in that context~\cite{Carrasco:2015xwa,Lubicz:2016xro}  might prove  useful for \NLDBD. 
On the EFT side,  one needs to compute full S-matrix elements, not potentials. 
As an illustration, and because the  $\pi \pi$  matrix element would probably be the first to be tested  on the lattice, 
we report the N$^2$LO S-matrix result for  $\pi^- (\vec{q})  \to \pi^+(\vec{q}) ee$, with $q^2 = m_\pi^2$:    
\be\label{eq44}
T_{\pi^- \to \pi^+ e e} =  T_{\rm lept} \times 2 \, F_\pi^2 \left[
1   + \frac{m_\pi^2}{(4 \pi F_\pi)^2}   \left(6 +  3  \log \frac{\mu^2}{m_\pi^2}  + \frac{5}{6} g_\nu^{\pi \pi} (\mu) \right) \right].
\ee%
In Eq. \eqref{eq44}, we set the electron four-momenta to zero.  
The S-matrix $T_{\pi^- \pi^- \to  e e}$ with generic kinematics is provided in Appendix \ref{AppA}.
The   S-matrix element for  $n \to p \pi^+  e e$ cannot be readily extracted from our matching
calculation, because we used off-shell ``potential" pions in the external legs of  Fig.~\ref{Fig1} (bottom panel).

Similarly,   $g_\nu^{NN}$  can be determined by matching the lattice calculation of $nn \to ppee$ 
to the chiral EFT one, with  a few  caveats: (i) the chiral EFT S-matrix elements requires a non-perturbative calculation, which we have not performed. 
(ii) Matching at N$^2$LO requires subtracting the ultrasoft contribution for the specific channel  $nn \to ppee$. 
In principle, all the ingredients ($E_n$ and  $\langle  pp | J_\mu | pn\rangle \langle np | J^\mu | nn \rangle$) 
to evaluate $T_{\rm usoft}$ can be computed in Lattice QCD, and a first step in this direction has been made in Ref.~\cite{Shanahan:2017bgi,Tiburzi:2017iux} 
in the context of the pionless EFT. (iii) Finally,  one needs to subtract the contributions from 
$g_{\nu}^{\pi \pi}$ and $g_{\nu}^{\pi N}$, or alternatively extract these from the $m_\pi$ dependence of the lattice 
$nn \to pp ee$ amplitude. 
\\

\noindent {\bf Pionless EFT:} 
To determine  the LO   (N$^2$LO)  couplings $g_\nu^{NN}$  ($g_{\nu,2}^{NN}$), 
one would  have to match  a Lattice QCD calculation of the $\Delta L=2$  $nn \to ppee$  scattering amplitude
to a full  LO  (N$^2$LO) calculation of the same amplitude in the  pionless EFT 
(or  the analogue  amplitude for the \NLDBD\ decay of the bound $nn$  state, relevant for heavy pion lattices ~\cite{Shanahan:2017bgi}).   
While obtaining the LO (N$^2$LO)  $nn \to ppee$  amplitude in pionless EFT is beyond the scope of this work,   we note that 
part of the LO amplitude is given in Eq.~\eqref{amp}.    
We also note that in performing the matching up to N$^2$LO one can ignore the contribution from the ultrasoft amplitude, 
which in pionless EFT  contributes beyond N$^2$LO  (see discussion in Sect.~\ref{sect:pionless}).
Should one need to evaluate $T_{\rm usoft}$,  the  input quantities  
$E_n$ and  $\langle  pp | J_\mu | p n\rangle \langle n p| J^\mu | nn \rangle$ 
can be  computed on the lattice~\cite{Shanahan:2017bgi,Tiburzi:2017iux}.

\section{Conclusions}

We have presented the first comprehensive effective theory analysis of \NLDBD\ induced by light Majorana-neutrino exchange, 
describing the physics from the scale $\Lambda_{\rm LNV}$ all the way down to the nuclear energy scale. 
The full  \NLDBD\  amplitude receives contributions from hard, soft, potential, and ultrasoft neutrino virtualities. 
Starting from the quark-level description, we have performed the matching to chiral EFT. 
In this context,  contributions from hard  modes are captured by local counterterms, 
while the  contributions from  soft and potential modes  can be explicitly evaluated and lead to appropriate
nuclear potentials -- insensitive to  properties of intermediate nuclear states  --   for which we have derived  LO and N$^2$LO expressions 
(see  Eqs.~\eqref{eq:Vnu},  \eqref{eq:Vnu0}, \eqref{eq:Vnu2}). 
We have identified new  contributions that  can not be captured by parameterizations of single nucleon form factors 
and by power-counting arguments are as large as  terms usually included in the \NLDBD\ amplitude.  
The contributions  from ultrasoft modes  appear at N$^2$LO and can be  displayed in terms of nuclear matrix 
elements of the weak current  and excitation energies of the intermediate odd-odd nuclei,   
that also  control  the    $2\nu \beta \beta$ amplitude (see Eq.~\eqref{eq:Tusoft}). 

In Section~\ref{sect:LECs} we discuss strategies to determine the low-energy constants  (LECs) that appear in the potentials. 
We have worked out  a connection  to the electromagnetic  LECs  encoding  the effect of hard  virtual photons in hadronic processes,  
that  can be obtained from model estimates, lattice QCD, and, at least  in principle, from data. 
We have also discussed a strategy to match directly  to 
 $\Delta I_z =2$ hadronic amplitudes  that could be  calculated in  Lattice QCD.

While the bulk of our discussion is based on the Weinberg version of chiral EFT, in Section~\ref{sect:pionless}
we also present the potential in pionless EFT to LO and N$^2$LO.  
We plan to study the consistency of the Weinberg power counting for \NLDBD\ decay in future work. 

In Section~\ref{sect:amplitude}  we have discussed  the  hierarchy of chiral EFT  contributions 
to the ``master formula" for the  \NLDBD\ amplitude,  Eq.\ \eqref{eq:Tnldbd}, 
describing their relation (when applicable)   to the  standard treatment of \NLDBD\ matrix elements in the literature.
We  advocate  that  many-body  calculations with existing methods~\cite{Horoi:2017gmj,Jiao:2017opc,Iwata:2016cxn,Barea:2015kwa,Hyvarinen:2015bda,Vaquero:2014dna,Yao:2014uta,Holt:2013tda,Simkovic:2013qiy,Menendez:2008jp,Haxton:2002kb,Haxton:1999vg},  as well as with methods under development~\cite{INTprogram},  
should be organized according to the EFT power counting scheme, 
isolating LO,  N$^2$LO, and ultrasoft contributions. 
Ideally, the neutrino potential derived here should be used with nuclear wavefunctions also based on chiral EFT and computed at next-to-leading order, or higher. This is particularly important when evaluating short range potentials. Benchmark calculations of double beta matrix elements of light nuclei \cite{Pastore:2017ofx} will help quantify the impact of the new N$^2$LO potential and ultimately assess the validity of the chiral framework.

\section*{Acknowledgements}
The idea for this work was formulated at a meeting of the Topical Collaboration on ``Nuclear Theory for Double-Beta Decay and Fundamental Symmetries'', supported the the DOE Office of Science.
VC  and  EM acknowledge support by the LDRD program at Los Alamos National Laboratory.
WD  acknowledges  support by the Dutch Organization for Scientific Research (NWO) 
through a RUBICON  grant. 
The work of AWL was supported in part by the Lawrence Berkeley National Laboratory (LBNL) LDRD program, the US DOE under contract DE-AC02-05CH11231 under which the Regents of the University of California manage and operate LBNL, the Office of Science Advanced Scientific Computing Research, Scientific Discovery through Advanced Computing (SciDAC) program under Award Number KB0301052 and by the DOE Early Career Research Program.
VC acknowledges partial support and hospitality of  the Mainz Institute for Theoretical Physics (MITP) 
during the completion of this work. 
We thank the  Institute for Nuclear Theory at the University of Washington for its hospitality and the Department of Energy 
for partial support during the completion of this work. 
We thank Martin Hoferichter for suggesting to us the connection between the electromagnetic low-energy constants 
and the ones needed for   \NLDBD, 
and Luigi Coraggio 
for informative exchanges.
We thank Javier Men\'endez for preliminary studies of the N$^2$LO potential in $^{136}$Xe.
We thank Bira van Kolck for several interesting discussions  and for comments on the manuscript.
We acknowledge discussions at various stages of this work with Joe Carlson, 
Zohreh Davoudi,  Jordy de Vries, 
Jon  Engel, Michael Graesser,   
Saori Pastore, and Martin Savage.

\appendix

\section{$\pi^- \pi^- \rightarrow e^- e^-$ scattering amplitude at N$^2$LO}\label{AppA}

In this Appendix we calculate the scattering amplitude for the process
$\pi^- (p_a) \pi^- (p_b) \rightarrow e^- (p_1) e^-(p_2)$, with on-shell pions $p_a^2 = p_b^2  = m_\pi^2$
and massless electrons $p_1^2 = p_2^2 = 0$, at N$^{2}$LO in Chiral Perturbation Theory. We introduce the Mandelstam variables
$s = (p_a + p_b)^2$, $t = (p_a - p_1)^2$, $u = (p_a - p_2)^2$.
The amplitude  $T_{\pi^- \pi^- \rightarrow e^- e^-}$ can be written as
\begin{equation}\label{t1}
T_{\pi^- \pi^- \rightarrow e^- e^-} = T_{\rm lept} \times 2 F_\pi^2 \mathcal S_{\pi\pi} +  4  G_F^2 V_{ud}^2  m_{\beta \beta}  \bar{u}_L (p_{1}) \sigma_{\mu\nu} C \bar{u}_L^T (p_{2})\, \mathcal A_{\pi\pi}^{\mu\nu},  
\end{equation}
where 
$T_{\rm lept} =  4  G_F^2 V_{ud}^2  m_{\beta \beta}  \bar{u}_L (p_{1}) C \bar{u}_L^T (p_{2})$, and
the antisymmetric leptonic structure  $\mathcal A_{\pi\pi}$ vanishes if $p^\mu_1 = p^\mu_2$.
Notice that we capture a subset of N$^2$LO corrections by normalizing the amplitude in terms of the physical pion decay constant $F_\pi$, rather than $F_0$.

For the amplitude $\mathcal S_{\pi\pi}$ we find
\begin{equation}
\mathcal S_{\pi\pi} = - \left[ \frac{1}{4} \left( \frac{1}{t} + \frac{1}{u}\right) (s -2 m_\pi^2) + \frac{1}{(4\pi F_\pi)^2} \left(\mathcal V_{\pi\pi} + \frac{s - 2 m_\pi^2}{2}\, \frac{5}{6} g_{\nu}^{\pi\pi}(\mu)\right) \right],
\end{equation}
where the loop contribution is given by
\begin{eqnarray}
\mathcal V_{\pi\pi} &=&
  \frac{3( s - 2 m_\pi^2)}{2}  \log \frac{\mu^2}{m_\pi^2}-\frac{\left(2 m_\pi^2-s\right)^2 \log ^2\left( -\frac{ 1+  \sqrt{1- \frac{4 m_\pi^2}{s}} }{1- \sqrt{ 1-\frac{4 m_\pi^2}{s}}  }\right)}{4 s}   \nonumber \\ 
& &-\frac{\left(m_\pi^2-t\right)  \log \left(1-\frac{t}{m_\pi^2}\right) \left( m_\pi^4 + 6 m_\pi^2 t +t (-s+t)\right)}{4 t^2} \nonumber \\
& & -\frac{\left(m_\pi^2-u\right) \log \left(1-\frac{u}{m_\pi^2}\right) \left(m_\pi^4 + 6 m_\pi^2 u +u (-s+u)\right)}{4 u^2} \nonumber \\
& & 
\nonumber  \\
 & & -\frac{ 6 m_\pi^4 (t + u) + 132 m_\pi^2\, t u +t u (-45 s+12 (t+u))}{24 t u}. 
\end{eqnarray}
At threshold, that is for  $s = 4 m_\pi^2$, $t = - m_\pi^2$, $u = -m_\pi^2$, we obtain 
\begin{eqnarray}
\mathcal S_{\pi\pi} &=&   1 - \frac{m_\pi^2}{(4\pi F_\pi)^2} \left( 3   \log \frac{\mu^2}{m_\pi^2} + \frac{7}{2}  + \frac{\pi^2}{4}  + \frac{5}{6} g_{\nu}^{\pi\pi}(\mu) \right).
\end{eqnarray}
At the kinematic point $s = 0$, $t = m_\pi^2$, $u = m_\pi^2$, $q^2 = m_\pi^2$,  which corresponds to the kinematics $\pi^{-}(q) \rightarrow \pi^{+}(q) e^-(0) e^-(0)$,
we recover Eq. \eqref{eq44}
\begin{eqnarray}
\mathcal S_{\pi\pi} &=&   1 +  \frac{m_\pi^2}{(4\pi F_\pi)^2} \left( 3 \log \frac{\mu^2}{m_\pi^2} + 6 + \frac{5}{6} g_{\nu}^{\pi\pi}(\mu)  \right).
\end{eqnarray}

\newpage
\bibliography{bibliography}

\end{document}